\documentclass[conference]{IEEEtran}
\IEEEoverridecommandlockouts
\usepackage{cite}
\usepackage{amsmath,amssymb,amsfonts}
\usepackage{algorithm} 
\usepackage{algpseudocode} 
\usepackage{graphicx}
\usepackage{hyperref}
\usepackage{adjustbox}
\usepackage{multirow}
\usepackage{booktabs}
\usepackage{textcomp}
\usepackage{xcolor}
\usepackage{placeins}
\usepackage{graphicx}
\usepackage{caption}
\usepackage{subcaption}
\usepackage{enumitem}

\def\BibTeX{{\rm B\kern-.05em{\sc i\kern-.025em b}\kern-.08em
    T\kern-.1667em\lower.7ex\hbox{E}\kern-.125emX}}
\usepackage{tikz}
\usetikzlibrary{shapes.geometric, arrows}

\begin{document}

\title{Topology-Aware Reinforcement Learning over Graphs for Resilient Power Distribution Networks}


\author{\IEEEauthorblockN{
\text {Roshni Anna Jacob}\IEEEauthorrefmark{1}, \textit{Member}, \textit{IEEE}, \text{Prithvi Poddar}\IEEEauthorrefmark{2},  \text{Jaidev Goel}\IEEEauthorrefmark{3}, \\ \text{Souma Chowdhury}\IEEEauthorrefmark{2}, \textit {Senior Member}, \textit {IEEE}, \text{Yulia R. Gel}\IEEEauthorrefmark{3}, and \text{Jie Zhang}\IEEEauthorrefmark{1}, \textit {Senior Member}, \textit {IEEE}
}
\IEEEauthorblockA{\IEEEauthorrefmark{1}The University of Texas at Dallas, Richardson, TX, USA\\
\IEEEauthorrefmark{2}University at Buffalo, Buffalo, NY, USA\\
\IEEEauthorrefmark{3}Virginia Tech, Blacksburg, VA, USA
}
Email: jiezhang@utdallas.edu
}
\maketitle

\begin{abstract}
Extreme weather events and cyberattacks can cause component failures and disrupt the operation of power distribution networks (DNs), during which reconfiguration and load shedding are often adopted for resilience enhancement. This study introduces a topology-aware graph reinforcement learning (RL) framework for outage management that embeds higher-order topological features of the DN into a graph-based RL model, enabling reconfiguration and load shedding to maximize energy supply while maintaining operational stability. Results on the modified IEEE 123-bus feeder across 300 diverse outage scenarios demonstrate that incorporating the topological data analysis (TDA) tool, persistence homology (PH), yields 9-18\% higher cumulative rewards, up to 6\% increase in power delivery, and 6-8\% fewer voltage violations compared to a baseline graph-RL model. These findings highlight the potential of integrating RL with TDA to enable self-healing in DNs, facilitating fast, adaptive, and automated restoration.
\end{abstract}


\section{Introduction}
With the growing penetration of distributed energy resources (DERs) and the increasing decentralization of the power grid, distribution networks (DNs) can be autonomously controlled to enhance resilience against disruptions caused by extreme events and cyber attacks~\cite{mahzarnia2020review}. Recent advancements in grid automation, particularly the deployment of remotely controllable switches and intelligent control devices, have led to the emergence of the self-healing concept in smart grids~\cite{refaat2018self}. This underscores the need for automatic, intelligent and fast-acting control of resources such as switches, loads, and DERs in response to operational disruptions within the DN.

Network reconfiguration and load shedding are two key operational strategies employed during disruptive events. Reconfiguration involves altering the network topology by controlling the open/closed status of switches~\cite{mishra2017comprehensive}. Through optimal and coordinated control of sectionalizing and tie switches, power can be rerouted to continue service to isolated sections intentionally forming self-sustaining islands~\cite{jacob2024real}.  
In parallel, load shedding may be necessary to alleviate overloads and ensure DN stability by maintaining the generation-demand balance~\cite{larik2018improved}. While these strategies have historically been part of utility operations, their implementation typically depends on static contingency plans that lack adaptability to changing system states. As disturbances in DNs can evolve within seconds, achieving active resilience necessitates rapid and adaptive decision-making~\cite{kandaperumal2020resilience}. Consequently, AI-driven control frameworks present a promising approach for real-time, data-informed operational response.

The outage management problem, involving network reconfiguration and load shedding, is traditionally formulated as a combinatorial nonlinear optimization problem and addressed using mixed-integer programming (MIP)~\cite{ma2018novel} or metaheuristic algorithms~\cite{liu2007skeleton}. However, the emergence of reinforcement learning (RL) has enabled model-free and adaptive decision-making in dynamic operating environments.
Studies such as~\cite{kundavcina2022solving,wang2021distribution} have applied RL for optimal network reconfiguration under normal operating conditions, primarily targeting loss minimization and voltage improvement. Methods based on deep Q-learning (DQN)~\cite{kundavcina2022solving} and its extensions, such as NoisyNet-enhanced exploration for improved parameter tuning~\cite{wang2021distribution}, have proven effective in normal settings since the feasible configuration space remains relatively tractable. However, under outage conditions, where the operating configuration evolves dynamically due to line failures and both islanding and rerouting are possible solutions, the state-action space expands dramatically, rendering action-value mapping infeasible. ~\cite{oh2025sequential} applied deep RL for dynamic sequential reconfiguration to improve reliability with renewable generation during normal conditions. Another class of studies, such as~\cite{ferreira2019reinforcement,zhao2021learning}, focused on RL models for service restoration under outage conditions. For example,~\cite{ferreira2019reinforcement} developed a Q-learning model and operated on aggregated feeder representations rather than network models with a larger state and action spaces. \cite{zhao2021learning} proposed a multi-agent RL framework with graph convolutional neural networks for sequential load restoration in blackstart scenarios, a related but distinct task from reconfiguration and load shedding. In our prior work~\cite{jacob2024real}, we developed an RL framework integrated with capsule-based graph convolutional networks for outage management in distribution networks, enabling reconfiguration and load shedding. This approach exploited the structural features of the problem through graph-based representations, providing the foundation for the topology-enhanced methods introduced in the present study.

This paper advances AI-driven outage management by developing a topological data analysis (TDA)-informed graph-based RL framework that serves as an intelligent resilience-support tool, enabling fast and adaptive reconfiguration and load shedding. Unlike conventional graph neural network (GNN) approaches, which do not explicitly capture the multi-resolution topological characteristics of DNs, the proposed framework integrates the TDA tool, persistence homology (PH), into the learning process. This PH-enhanced graph RL framework leverages higher-order topological information to improve decision-making and provides a principled means of quantifying grid resilience through topological descriptors.

The remainder of the paper is organized as follows. Section~\ref{Problem Formulation} presents the problem formulation including topological data analysis. Section~\ref{RL over Graphs} describes the RL framework along with TDA integration. Section~\ref{Results} presents the results and discussion, and Section~\ref{Conclusion} concludes the paper.

\section{Problem Formulation}
\label{Problem Formulation}

\subsection{Overview}
\vspace{-0.2em}
The outage management involves reconfiguring the network to establish alternate paths for power flow from the substation to disconnected segments or leveraging grid-forming DERs to support islanded sections. Selective load shedding is also employed to maintain the load-generation balance within reconfigured network segments. A comprehensive description of the problem formulation and baseline approaches can be found in our previous work~\cite{jacob2024real}.

\subsection{Markov Decision Process (MDP) over graphs}

The DN can be mathematically represented as a graph $\mathcal{G}=(\mathcal{N}, \mathcal{E})$ where $\mathcal{N}$ is the set of nodes and $\mathcal{E}$ is the set of edges. The buses in the DN form the nodes of the graph, and an edge $e_{ij}\in\mathcal{E}$ exists if the buses corresponding to nodes $i$ and $j$ are physically connected. $\mathbf{A}\in\mathbb{R}^{|\mathcal{N}|\times |\mathcal{N}|}$ represents the weighted adjacency matrix of the graph.
The resilience improvement problem in DN is formulated as an MDP over this graph-based representation, formally defined as ($\mathcal{S}, \mathcal{A}, R, P_{tr}$). The state space ($\mathcal{S}$) captures DN features, including bus voltages, branch flows, energy supplied, network configuration, power flow violations, and switch outage masks. The actions ($\mathcal{A}$) comprise discrete control decisions for line and load switching. The reward ($R$) quantifies resilience objectives by maximizing, energy supplied in the network while penalizing voltage violations and power-flow convergence failures. It is defined as:
    \begin{equation}
        R(s,a) = 
        \begin{cases} 
            E_{\text{supp}} - V_{\text{viol}}, & \text{if} \ C_{\text{viol}} = 0, \\ 
            -1, & \text{if} \ C_{\text{viol}} = 1. 
        \end{cases}
    \label{rewardEq}
    \end{equation}
Here, $E_{\text{supp}}$ denotes the total energy supplied measured as per unit of the total demand in the network. The convergence flag $C_{\text{viol}}$ is set to 1 in the presence of power flow convergence issues, invalid observations, or exceptions in the OpenDSS simulations, resulting in a penalty of $-1$. Additionally, $V_{\text{viol}}$ represents the aggregate voltage violation across the network measured as:
\begin{equation}
\begin{split}
V_{\text{viol}} =
\frac{1}{3|\mathcal{N}|}
\sum_{i \in \mathcal{N}} \sum_{j \in \phi_i}
\Big[ &
\max\{V_j^i - V^{\max},\, 0\} \\[-12pt]
& + \max\{V^{\min} - V_j^i,\, 0\}
\Big]
\end{split}
\label{eq:Vviol_combined}
\end{equation}
Here, $V_j^i$ is the voltage magnitude at phase $j$ of node $i$, and $\phi$ denotes the set of active phases at that node.
 The transition dynamics ($P_{tr}$) are implicitly learned through interactions between the RL agent and the simulation environment. 
The RL environment integrates OpenDSS-based power flow simulations with graph representations, and strict radiality is not explicitly enforced. This is because radial restoration topologies emerge as optimal due to the restoration objective and underlying distribution-network physics, while explicit radiality penalties were avoided as they degraded learning convergence without improving restoration performance. Additionally, emergency restoration and islanded microgrid operation do not necessarily follow normal protection-driven radial operation assumptions. Each node $i\in\mathcal{N}$ is represented with a feature vector denoting the 3-phase voltages measured at the node. The feature vector of all the nodes collectively is represented in learning architecture as a node feature matrix $\mathbf{V}\in\mathbb{R}^{|\mathcal{N}|\times 3}$. State information not encoded in the graph nodes is treated as contextual input to the RL agent.

\subsection{Topological Data Analysis}
TDA, and in particular its subfield PH, provides a systematic framework for extracting multi-scale structural information from data. The core idea is to analyze a dataset $\mathbb{X}$ across varying resolutions and to track the appearance and disappearance of topological patterns as the resolution threshold changes monotonically.
Let $\alpha>0$ denote the threshold. A proximity graph $G_{\alpha}(\mathbb{X})$ is constructed by connecting two data points whenever their pairwise distance is less than $\alpha$. By considering a monotonic sequence of thresholds $\alpha_1 < \alpha_2 < \ldots < \alpha_n$, a nested sequence of proximity graphs $G_{\alpha_1}(\mathbb{X}) \subseteq G_{\alpha_2}(\mathbb{X}) \subseteq \ldots \subseteq G_{\alpha_n}(\mathbb{X})$ is obtained forming a graph filtration.

To study the evolution of topological summaries, such as the number of connected components, loops, and voids, each proximity graph is associated with an abstract simplicial complex $k$, producing a corresponding filtration of complexes:
$k_{\alpha_1}(\mathbb{X}) \subseteq k_{\alpha_2}(\mathbb{X}) \subseteq \ldots \subseteq k_{\alpha_n}(\mathbb{X})$.
Vietoris-Rips complex is one of the most popular choices for $k$ due to its computational advantages.
Topological features that persist across a wide range of thresholds $(\alpha_1 < \alpha_2 < \ldots < \alpha_n)$ are typically indicative of the intrinsic structural organization of $\mathbb{X}$, while those with short lifespans are often regarded as topological noise. For each topological feature $\delta$, the parameters $b_{\delta}$ and $d_{\delta}$ denote the indices at which the feature appears and disappears, respectively. The pair $(b_{\delta}, d_{\delta})$ thus represents the birth and death of feature $\delta$, with $(d_{\delta} - b_{\delta})$ corresponding to its persistence or lifespan. Collectively, these $(b_{\delta}, d_{\delta})$ pairs form a multiset in $\mathbb{R}^2$, known as the persistence diagram (PD), which encodes the multi-scale topological signatures of the data.

\section{Reinforcement Learning over Graphs}\label{RL over Graphs}
This study employs Proximal Policy Optimization (PPO) \cite{schulman2017proximalpolicyoptimizationalgorithms} algorithm, an RL method, to address the DN outage management problem. The policy network is designed as a topology-aware graph neural network (GNN) that captures spatial dependencies among buses and branches.
\begin{figure}[tbp]
  \centering  
    \includegraphics[width=0.75\columnwidth]{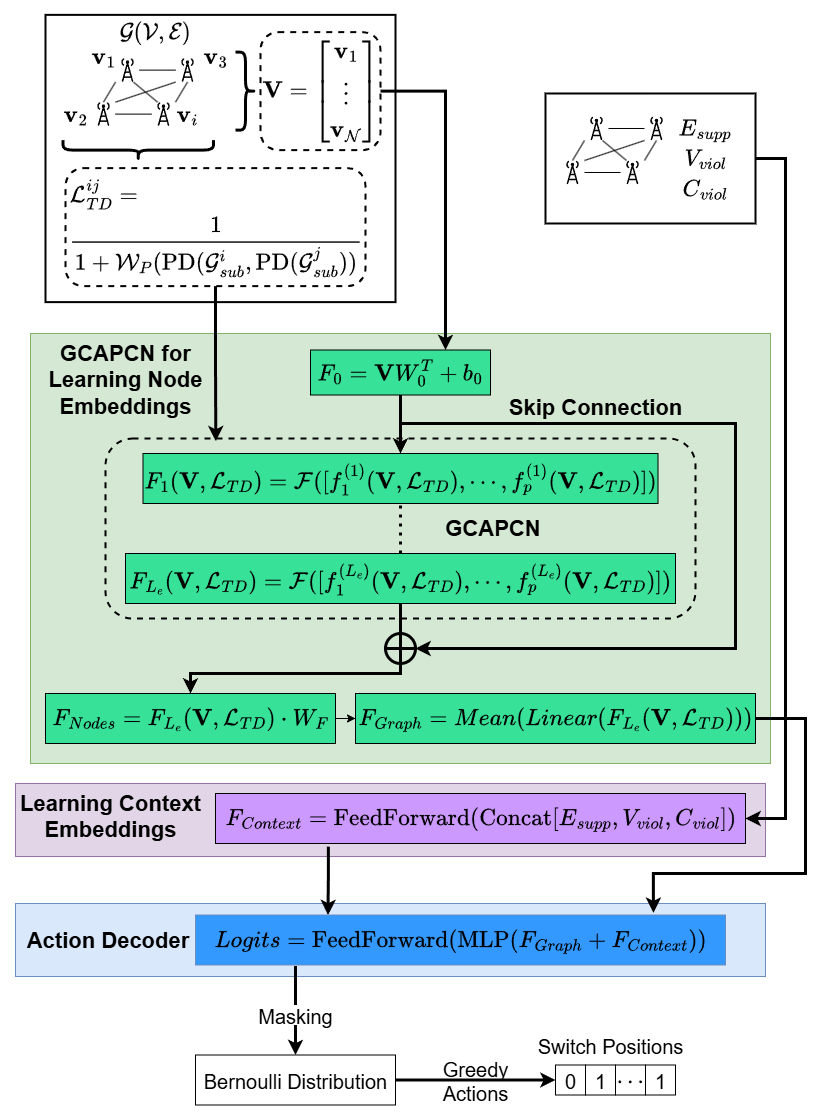}
    \caption{Architecture of the policy network (PH-GCAPCN).}
    \label{fig:architecture}
    \vspace{-2em}
\end{figure}

\subsection{Environment}
The environment is developed using the open-source distribution system simulator (OpenDSS). DERs are modeled using the built-in Generator and PVSystem classes in OpenDSS. Sectionalizing and tie switches are predefined and controlled using the switch control function. Load shedding or pick up is performed at the primary feeder level, where aggregated loads at distribution transformer nodes can be selectively enabled or disabled. Power flow computation, execution of control actions, and extraction of system observations are implemented in Python using OpenDSSDirect.

\vspace{-0.53em}
\subsection{Learning Architecture}
To effectively capture the topological interdependence of the state variables, a GNN-based policy architecture is employed. The overall architecture is illustrated in Fig.~\ref{fig:architecture}.
The policy network comprises three main components: 
\vspace{-0.2em}
\begin{enumerate}[leftmargin=1em]
\item A graph capsule convolutional neural network (GCAPCN)~\cite{verma2018graphcapsuleconvolutionalneural} that takes the DN graph $\mathcal{G}$ as input and learns higher-dimensional node embeddings for downstream actions. A linearly transformed node feature vector ($F_0$) is passed through a series of capsule-based graph convolution layers, where the $l^{th}$ layer computes a node feature matrix using:
\vspace{-0.5em}
\begin{equation}
    F_l(\mathbf{V},L)=\mathcal{F}_{-2}([f_1^{(l)}(\mathbf{V},L), f_2^{(l)}(\mathbf{V},L),\cdots,f_p^{(l)}(\mathbf{V},L)]),
\end{equation}
where $f_i^{(l)}$ $(\mathbf{V},L)\in\mathbb{R}^{|\mathcal{N}|\times h_l}$ denotes the $i^{\text{th}}$ capsule, $p$ is the maximum statistical moment order, and $\mathcal{F}_{-2}(.)$ flattens the last two dimensions of a tensor. Each capsule is computed via a polynomial graph convolution operator:
\vspace{-0.5em}
\begin{equation}
\begin{split}
    &f_i^{(l)}(\mathbf{V},L)= \\[-6pt] &\sigma \left( \sum_{k=0}^{K} {L}^k \left( \underbrace{ F_{(l-1)} ({\mathbf{V}}, {L}) \odot \dots \odot F_{(l-1)} ({\mathbf{V}}, {L}) }_{i \text{ times}} \right) {W}_{ik}^{(l)} \right)
\end{split}
\end{equation}
where $F_{(l-1)} ({\mathbf{V}}, {L})\in\mathbb{R}^{|\mathcal{N}|\times h_{l-1}p}$ is the output from the $l-1$ layer, $W_{ik}^{(l)}\in \mathbb{R}^{h_{l-1}p\times h_l}$ is a learnable weight matrix, and $K$ is the degree of the convolutional filter.
For computing the final node embeddings $F_{Nodes}$, the output of the last layer (of dimension $h_{Le}$) is passed through a linear transformation. Finally,the graph-level embedding is computed by aggregating node embeddings as:
\vspace{-0.5em}
\begin{equation}
    F_{graph}=\text{Mean}(W_{g2}\cdot(W_{g1}\cdot F_{Nodes})),
\end{equation}
where $W_{g1}\in\mathbb{R}^{h_{Le}\times|\mathcal{N}|}$, $W_{g2}\in\mathbb{R}^{h_{Le}\times h_{Le}}$, and $F_{graph}\in\mathbb{R}^{h_{Le}}$.
\item  A feed-forward network that processes contextual information such as the total energy supplied in the network, voltage violations and the branch power flows. The feature vector, formed by concatenating total energy supplied ($E_{supp}$), voltage violations ($V_{viol}$), and branch flows ($b_e$), is processed as:
\vspace{-0.5em}
\begin{equation}
    F_{context}=\text{Feedforward}(\text{Concat}[E_{supp},V_{viol},b_e]).
\end{equation}
\item An action decoder that integrates the node embeddings and contextual information to determine the next control action of the RL agent. The final graph embedding ($F_{\text{graph}}$) and context vector ($F_{\text{context}}$) are combined and processed through an MLP followed by a feedforward layer to compute the action logits:
\vspace{-0.5em}
\begin{equation}
    Logits=\text{Feedforward}(\text{MLP}(F_{graph}+F_{context}).
\end{equation}
\end{enumerate}
The logits of masked switches are set to $-\infty$, after which a Bernoulli distribution over all actions is derived using sigmoid-normalized probabilities; the final switching action follows a greedy policy, setting a switch to ON if its mean activation exceeds 0.5.

To capture structural information beyond pairwise relationships within DNs, tools from PH are employed to enhance the GCAPCN architecture through topologically-informed edge reweighting. The proposed approach involves: (1) Local neighborhood extraction: for each node $i \in \mathcal{N}$, a $k$-hop neighborhood subgraph $\mathcal{G}_i = (\mathcal{N}_i, \mathcal{E}_i)$ is constructed, where $\mathcal{N}_i$ includes all nodes within a spatial distance $k$ of node $i$; (2) Persistence diagram computation: for each subgraph $\mathcal{G}_i$, a persistence diagram $PD_i = {(b_j, d_j)}$ is computed, encoding the birth and death times of topological features across multiple scales; (3) Topological edge reweighting: The binary adjacency matrix $\mathbf{A}$ is replaced with a topologically aware matrix $\mathbf{A}_{PH}$, where edge weights are defined as $\mathcal{L}_{TD}^{i,j} = 1/(1 + \mathcal{W}_2(PD_i, PD_j))$, and $\mathcal{W}_2$ denotes the 2-Wasserstein distance between PDs. This reweighting assigns higher connectivity weights to node pairs with similar topological signatures, enabling the GNN to aggregate information from nodes that exhibit comparable structural roles in network restoration, independent of their physical proximity.

 \begin{figure}[tbp]
  \centering  
  \begin{subfigure}[h]{0.48\linewidth}
    \centering
    \includegraphics[width=1.05\columnwidth]{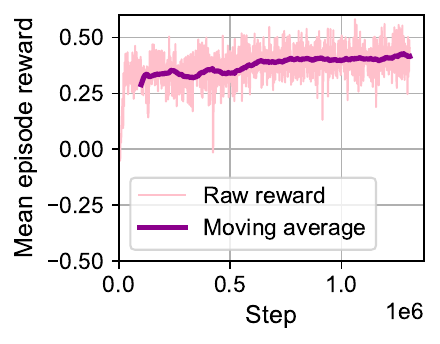}
    \caption{PH-GCAPCN reward curve}
    \label{fig:PHCAPAM}
  \end{subfigure}
  \hfill
  \begin{subfigure}[h]{0.48\linewidth}
    \centering
    \includegraphics[width=1.05\columnwidth]{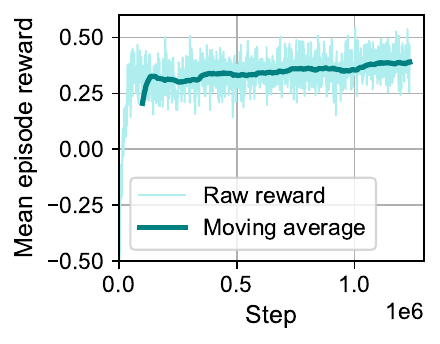}
    \caption{GCAPCN reward curve}
    \label{fig:CAPAM}
  \end{subfigure} 
  \caption{Training performance for the proposed RL model based on PH-GCAPCN and the baseline RL model using GCAPCN.}
  \label{fig:reward_comparison}
  \vspace{-1.8em}
\end{figure}

\section{Results}
\label{Results}
\subsection{Test Network}
The topology-aware RL model (PH-GCAPCN) and the baseline RL model (GCAPCN) were validated using a modified IEEE 123-bus distribution test feeder. The DN circuit within the environment was modified by adding 13 sectionalizing switches and 9 tie switches, each with assigned switching controls. Grid-forming DERs rated at 250 kW were installed at buses 39, 46, 71, 75, 79, 96, and 108, while grid-feeding DERs rated at 80 kW were placed at buses 11, 33, 56, 82, 91, and 104. Detailed network information is presented in~\cite{jacob2024real}.                                                                 
\subsection{Experiment Details}
The training dataset comprised 10,000 unique outage scenarios designed to represent diverse network disruption conditions. Each scenario was generated by selecting an outage center from 25 spatially distributed nodes and extracting a subgraph $\mathcal{G}_{sub}$ with radius $r \sim U\left(1, \tfrac{1}{3}\mathrm{diam}(\mathcal{G})\right)$. An outage severity parameter $s \sim U(0, 0.3)$ determined the number of disconnected lines $k = \max\left(1, \lceil s |\mathcal{E}_{sub}| \rceil\right)$, chosen uniformly at random within the subgraph. This process emulates localized, cascading outages and provided the pool of scenarios for model training. All training was performed on a university-managed HPC cluster using SLURM scheduling. Jobs ran on the A30 GPU partition (NVIDIA A30, 24 GB VRAM), with one GPU per job. The GCAPCN model completed training within a single 48-hour job, whereas the PH-GCAPCN model required extended training executed as five sequential 48-hour jobs, each resuming from the previous checkpoint to meet wall-time limits.

During testing, different 100 disjoint outage scenarios were generated and validated in OpenDSS to ensure power-flow convergence and valid system states, forming fixed and reproducible test sets for performance evaluation. Inference was performed on a Windows 11 Enterprise system (Intel Core i7-1365U, 16 GB RAM), with average computation times of 0.01887 ± 0.0042 s for PH-GCAPCN and 0.0098 ± 0.0017 s for GCAPCN over 10 runs; these values are not directly comparable to HPC training runtimes due to hardware differences.
The implementation of the proposed model and along with the testing scripts is available at~\url{https://github.com/RoshniAnna/RL-PHGCAPCN-Grid-Resilience}

\begin{table*}[htbp]
\centering
\caption{Performance of PH-GCAPCN and GCAPCN RL models across three test samples. Values are reported as mean$\pm$std over 100 distinct out-of-sample scenarios.}
\renewcommand{\arraystretch}{1.2}
\setlength{\tabcolsep}{6pt}
\vspace{-0.5em}
\begin{tabular}{l|c|c|c|c|c|c}
\hline
Metric & \multicolumn{2}{c|}{Test Set 1} & \multicolumn{2}{c|}{Test Set 2} & \multicolumn{2}{c}{Test Set 3} \\
\cline{2-7}
 & \textbf{PH-GCAPCN} & GCAPCN & \textbf{PH-GCAPCN} & GCAPCN & \textbf{PH-GCAPCN} & GCAPCN \\
\hline
Reward ($R$) &  \textbf{0.4626$\pm$0.5757}  &  0.3923$\pm$0.5692 & \textbf{0.5132$\pm$0.5540} & 0.4710$\pm$0.5407 & \textbf{0.5538$\pm$0.4842} &  0.5010$\pm$0.4659 \\
Energy Supplied ($E_{\text{supp}}$) & \textbf{0.6760$\pm$0.3433}  & 0.6374$\pm$0.3265 & \textbf{0.7119$\pm$0.3234}
 & 0.6821$\pm$0.3129 & \textbf{0.7302$\pm$0.2913} &  0.6921$\pm$0.2769 \\
Voltage Violation ($V_{\text{viol}}$) &  \textbf{0.2134$\pm$0.2330}
 & 0.2282$\pm$0.2276 & \textbf{0.1932$\pm$0.2189}
 & 0.2057$\pm$0.2166 & \textbf{0.1763$\pm$0.1942} &  0.1911$\pm$0.1909 \\
\hline
\end{tabular}
\label{Tab:ModelResults}
\vspace{-2pt}
\end{table*}

\begin{table}[htbp!]
\centering
\caption{Paired t-test $p$-values comparing PH-GCAPCN and GCAPCN over 100 matched outage scenarios per test sample.}
\vspace{-0.3em}
\begin{tabular}{lccc}
\hline
\textbf{Test Set} & \textbf{Reward} & \textbf{Energy Supplied} & \textbf{Voltage Violation} \\
\hline
1 & $6.83\times10^{-4}{}^{***}$ & $1.26\times10^{-8}{}^{***}$ & $9.42\times10^{-4}{}^{***}$ \\
2 & $3.82\times10^{-5}{}^{***}$ & $3.75\times10^{-9}{}^{***}$ & $2.73\times10^{-5}{}^{***}$ \\
3 & $2.15\times10^{-7}{}^{***}$ & $7.39\times10^{-8}{}^{***}$ & $9.04\times10^{-5}{}^{***}$ \\
\hline
\end{tabular}
\smallskip
\raggedright
\centering
\footnotesize{\textit{Significance codes:} $^{***}p < 0.001$, $^{**}p < 0.01$, $^{*}p < 0.05$.}
\label{Tab:PairedTtest}
\vspace{-2.5em}
\end{table}

\begin{figure*}[h!]
\centering
\begin{minipage}{0.32\linewidth}
\centering
  \includegraphics[width=0.9\linewidth]{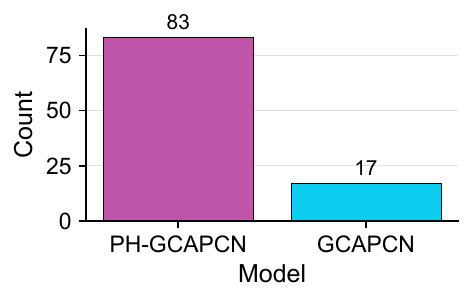}\\[-0.4ex]
  \small\textbf{(a)} Win rate for test set 1
\end{minipage}\hfill
\begin{minipage}{0.32\linewidth}
\centering
  \includegraphics[width=0.9\linewidth]{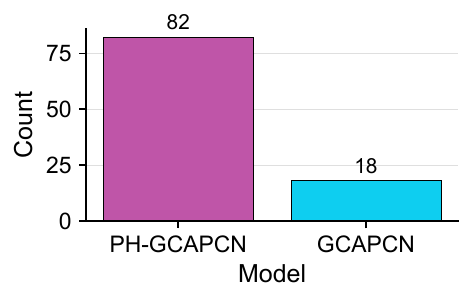}\\[-0.4ex]
  \small\textbf{(b)} Win rate for test set 2
\end{minipage}\hfill
\begin{minipage}{0.32\linewidth}
\centering
  \includegraphics[width=0.9\linewidth]{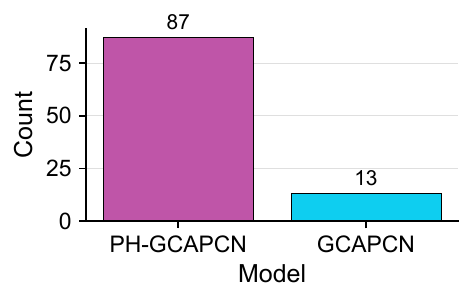}\\[-0.4ex]
  \small\textbf{(c)} Win rate for test set 3
\end{minipage}

\vspace{0.6ex}

\begin{minipage}{0.32\linewidth}
\centering
  \includegraphics[width=0.9\linewidth]{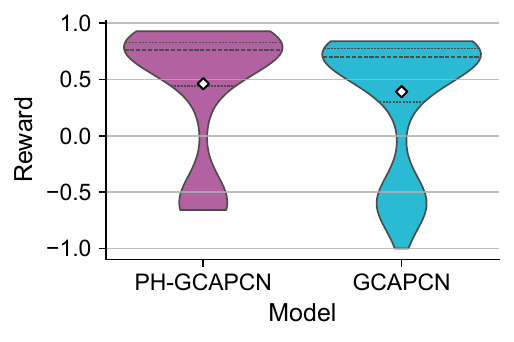}\\[-0.4ex]
  \small\textbf{(d)} Reward distribution for test set 1
\end{minipage}\hfill
\begin{minipage}{0.32\linewidth}
\centering
  \includegraphics[width=0.9\linewidth]{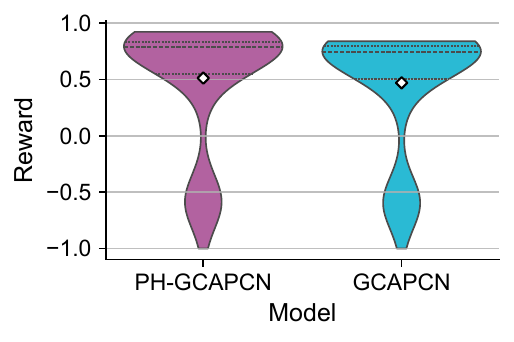}\\[-0.4ex]
  \small\textbf{(e)} Reward distribution for test set 2
\end{minipage}\hfill
\begin{minipage}{0.32\linewidth}
\centering
  \includegraphics[width=0.9\linewidth]{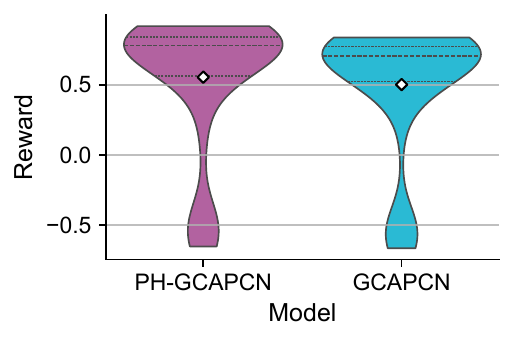}\\[-0.4ex]
  \small\textbf{(f)} Reward distribution for test set 3
\end{minipage}

\caption{Test performance comparison of PH-GCAPCN and GCAPCN models}
\label{fig:winrate-violin}
\vspace{-1em}
\end{figure*}


\subsection{Model Comparison}

To assess the impact of topological information, the PH-GCAPCN RL model is compared with the baseline GCAPCN model, both trained under identical conditions using randomly sampled initial states from a common outage scenario set.
Figure~\ref{fig:reward_comparison} presents the training reward trajectories for both models, with moving-average trends (darker curves) overlaid on the raw reward curves (lighter lines). The PH-GCAPCN model (Fig.~\ref{fig:PHCAPAM}) shows a rapid rise in mean episode reward within the first 100k steps, followed by steady improvement until around 500k steps, after which convergence is observed. In contrast, the GCAPCN model exhibits slower reward growth. At convergence, PH-GCAPCN achieves a moving-average reward about 0.04 p.u. higher than GCAPCN, with peaks up to 0.2 p.u. during training, indicating that topology-informed learning enhances generalization and stabilizes policy updates.

Table~\ref{Tab:ModelResults} summarizes the mean and standard deviation for proposed and baseline model performance over 100 distinct out-of-sample scenarios for three independent test samples. Besides reward, energy supplied (total energy served in per units; higher indicates greater resilience), and  voltage violation (averaged per-unit deviation outside [0.95, 1.05]; lower indicates better operational stability) are also reported. Across all test samples, PH-GCAPCN consistently outperforms GCAPCN on all metrics, achieving 9-18\% higher rewards, 4-6\% more energy supplied, and 6-8\% fewer voltage violations. Comparable standard deviations indicate a true performance gain rather than increased variance. Testing on disjoint out-of-sample scenarios further confirms the model’s generalizability, and Table~\ref{Tab:PairedTtest} reports the statistical significance of these improvements.

Figure~\ref{fig:winrate-violin} summarizes the scenario-level comparison between PH-GCAPCN and GCAPCN during testing. The top row shows the win counts, where a “win” equals a higher reward for a given scenario between the two models. Across the three disjoint test samples, the RL model based on PH-GCAPCN wins 83/100, 82/100, and 87/100 scenarios, respectively, indicating model superiority. The bottom row in Fig.~\ref{fig:winrate-violin} presents violin plots of the reward distributions across the test scenarios. The violin width indicates density, the horizontal lines mark the quartiles (the middle line is the median), and the white diamonds mark the mean. Compared to the GCAPCN baseline, the PH-GCAPCN distribution is shifted upward, with higher mean and median values. This indicates that PH-GCAPCN delivers improved rewards across the majority of scenarios, rather than depending on a few outlier gains.

\vspace{-0.5em}
\section{Conclusion}
\label{Conclusion}
This paper presents a topology-aware reinforcement learning framework for outage mitigation in power distribution networks, integrating persistent homology into a graph-based learning architecture to capture higher-order topological features and enhance switching decisions. Trained and validated on the modified 123-bus environment with switches and distributed energy resources, the proposed method outperforms the baseline graph RL model, achieving higher rewards, increased energy delivery, and reduced voltage violations. Future work will explore decision interpretability and extend the approach to generator dispatch under outage conditions.

\vspace{-0.5em}
\section*{Acknowledgment}
This work was partially supported by the Department of the Navy, Office of Naval Research under ONR award number N00014-21-1-2530 and National Science Foundation under Award 2229417. 
\vspace{-1em}
\FloatBarrier
\bibliographystyle{IEEEtran}
\bibliography{reference}

@misc{schulman2017proximalpolicyoptimizationalgorithms,
      title={Proximal Policy Optimization Algorithms}, 
      author={John Schulman and Filip Wolski and Prafulla Dhariwal and Alec Radford and Oleg Klimov},
      year={2017},
      eprint={1707.06347},
      archivePrefix={arXiv},
      primaryClass={cs.LG},
      url={}, 
}

@misc{verma2018graphcapsuleconvolutionalneural,
      title={Graph Capsule Convolutional Neural Networks}, 
      author={Saurabh Verma and Zhi-Li Zhang},
      year={2018},
      eprint={1805.08090},
      archivePrefix={arXiv},
      primaryClass={stat.ML},
      url={https://arxiv.org/abs/1805.08090}, 
}

@article{kundavcina2022solving,
  title={Solving dynamic distribution network reconfiguration using deep reinforcement learning},
  author={Kunda{\v{c}}ina, Ognjen B and Vidovi{\'c}, Predrag M and Petkovi{\'c}, Milan R},
  journal={Electrical Engineering},
  volume={104},
  number={3},
  pages={1487--1501},
  year={2022},
  publisher={Springer}
}

@article{wang2021distribution,
  title={Distribution network reconfiguration based on NoisyNet deep Q-learning network},
  author={Wang, Beibei and Zhu, Hong and Xu, Honghua and Bao, Yuqing and Di, Huifang},
  journal={IEEE Access},
  volume={9},
  pages={90358--90365},
  year={2021},
  publisher={IEEE}
}

@article{oh2025sequential,
  title={Sequential Control of Individual Switches for Real-time Distribution Network Reconfiguration Using Deep Reinforcement Learning},
  author={Oh, Jae-Young and Oh, Seok Hwa and Lee, Gyu-Sub and Yoon, Yong Tae and Jo, Seungchan},
  journal={IEEE Transactions on Smart Grid},
  year={2025},
  publisher={IEEE}
}

@article{ferreira2019reinforcement,
  title={A reinforcement learning approach to solve service restoration and load management simultaneously for distribution networks},
  author={Ferreira, Lucas Roberto and Aoki, Alexandre Rasi and Lambert-Torres, Germano},
  journal={IEEE Access},
  volume={7},
  pages={145978--145987},
  year={2019},
  publisher={IEEE}
}

@article{zhao2021learning,
  title={Learning sequential distribution system restoration via graph-reinforcement learning},
  author={Zhao, Tianqiao and Wang, Jianhui},
  journal={IEEE Transactions on Power Systems},
  volume={37},
  number={2},
  pages={1601--1611},
  year={2021},
  publisher={IEEE}
}

@article{jacob2024real,
  title={Real-time outage management in active distribution networks using reinforcement learning over graphs},
  author={Jacob, Roshni Anna and Paul, Steve and Chowdhury, Souma and Gel, Yulia R and Zhang, Jie},
  journal={Nature Communications},
  volume={15},
  number={1},
  pages={4766},
  year={2024},
  publisher={Nature Publishing Group UK London}
}

@article{mahzarnia2020review,
  title={A review of the measures to enhance power systems resilience},
  author={Mahzarnia, Maedeh and Moghaddam, Mohsen Parsa and Baboli, Payam Teimourzadeh and Siano, Pierluigi},
  journal={IEEE Systems Journal},
  volume={14},
  number={3},
  pages={4059--4070},
  year={2020},
  publisher={IEEE}
}

@inproceedings{refaat2018self,
  title={Self-Healing control strategy; Challenges and opportunities for distribution systems in smart grid},
  author={Refaat, Shady S and Mohamed, Amira and Kakosimos, Panagiotis},
  booktitle={2018 IEEE 12th International conference on compatibility, power electronics and power engineering (CPE-POWERENG 2018)},
  pages={1--6},
  year={2018},
  organization={IEEE}
}

@article{mishra2017comprehensive,
  title={A comprehensive review on power distribution network reconfiguration},
  author={Mishra, Sivkumar and Das, Debapriya and Paul, Subrata},
  journal={Energy Systems},
  volume={8},
  number={2},
  pages={227--284},
  year={2017},
  publisher={Springer}
}

@article{larik2018improved,
  title={An improved algorithm for optimal load shedding in power systems},
  author={Larik, Raja Masood and Mustafa, Mohd Wazir and Aman, Muhammad Naveed and Jumani, Touqeer Ahmed and Sajid, Suhaib and Panjwani, Manoj Kumar},
  journal={Energies},
  volume={11},
  number={7},
  pages={1808},
  year={2018},
  publisher={MDPI}
}

@article{liu2007skeleton,
  title={Skeleton-network reconfiguration based on topological characteristics of scale-free networks and discrete particle swarm optimization},
  author={Liu, Yan and Gu, Xueping},
  journal={IEEE Transactions on Power Systems},
  volume={22},
  number={3},
  pages={1267--1274},
  year={2007},
  publisher={IEEE}
}

@inproceedings{ma2018novel,
  title={A novel MILP formulation for fault isolation and network reconfiguration in active distribution systems},
  author={Ma, Shanshan and Li, Shiyang and Wang, Zhaoyu and Arif, Anmar and Ma, Kang},
  booktitle={2018 IEEE Power \& Energy Society General Meeting (PESGM)},
  pages={1--5},
  year={2018},
  organization={IEEE}
}

@article{kandaperumal2020resilience,
  title={Resilience of the electric distribution systems: concepts, classification, assessment, challenges, and research needs},
  author={Kandaperumal, Gowtham and Srivastava, Anurag K},
  journal={IET Smart Grid},
  volume={3},
  number={2},
  pages={133--143},
  year={2020},
  publisher={Wiley Online Library}
}
\end{document}